\documentstyle[aps,prl,twocolumn,epsfig]{revtex}
\draft
\begin{document}
\title{
Kondo time scales for quantum dots --- response to pulsed bias potentials}
\author{Martin Plihal and David C. Langreth}
\address{Center for Materials Theory,
Department of Physics and Astronomy,
Rutgers University, Piscataway, NJ 08854-8019}
\author{Peter Nordlander}
\address{
Department of Physics and Rice Quantum Institute,
Rice University, Houston, Texas 77251-1892 }

\maketitle
\thispagestyle{empty}
\begin{abstract}
The response of a quantum dot in the Kondo regime to rectangular
pulsed bias potentials of various strengths and durations
is studied theoretically. 
It is found that the rise time is faster than the fall time,
and also faster than time scales
normally associated with the Kondo problem. For larger values of
the pulsed bias,
 one can induce dramatic oscillations in the induced current
with a frequency approximating the splitting between the Kondo
peaks that would
be present in 
steady state.
The effect persists in the total charge transported per pulse,
which
should facilitate the
experimental observation
of the phenomenon.
\end{abstract}

\narrowtext

The theoretical predictions \cite{theory} of consequences of the Kondo effect
for the steady state conduction through 
quantum dots began a decade ago.
At low temperatures, a narrow resonance in the dot density
of states can form at the Fermi level, leading to a large
 enhancement of the dot's conductance, which is strongly dependent
on temperature, bias, and magnetic field.
Many of these effects have been recently observed by a
set of beautiful experiments by several groups
\cite{experiment}. These successes, supplemented by the
anticipation  that time dependent experiments \cite{acreview} are
not far behind, have spurred a number of theoretical groups
\cite{drivenkondo} to consider the effects expected when
sinusoidal biases or gate potentials are applied. Surprisingly,
the application of steps or pulses, which can provide
a less ambiguous measure of time scales, have not been considered
until very recently \cite{kondotime};  the latter work considered the
time dependent change in linear response conductance when
a  stepped potential was applied to a gate, thereby shifting the
dot into the Kondo regime. In the present work we consider the
response of a dot already in the Kondo regime to a sudden change of
the bias potential across the dot. We show that the physics is
qualitatively different from the latter case \cite{kondotime}, leading
to a different range of characteristic times and  physical phenomena. 

While Ref.~\onlinecite{kondotime} studied the time scale for the system
to go from one equilibrium configuration to another, we study here
the time scale to go from an equilibrium configuration to a non-equilibrium
one; and then back to equilibrium again. We find that 
these latter two times scales are 
very different from each other, the first being much shorter -- and
also much shorter than the time scale of \onlinecite{kondotime}. Furthermore,
if one applies a rectangular bias pulse  large enough to split the 
Kondo resonance, then there
appear current oscillations \cite{coleman} at a frequency  characterizing
the splitting between the Kondo peaks.
  We show that these current oscillations cause
oscillations in the charge transported through the dot as a function of
pulse length, thus providing a clear experimental signature.  The
damping of the oscillations provides an additional time scale which also
can be measured directly.

\begin{figure}
\centerline{\epsfxsize=0.46\textwidth
\epsfbox{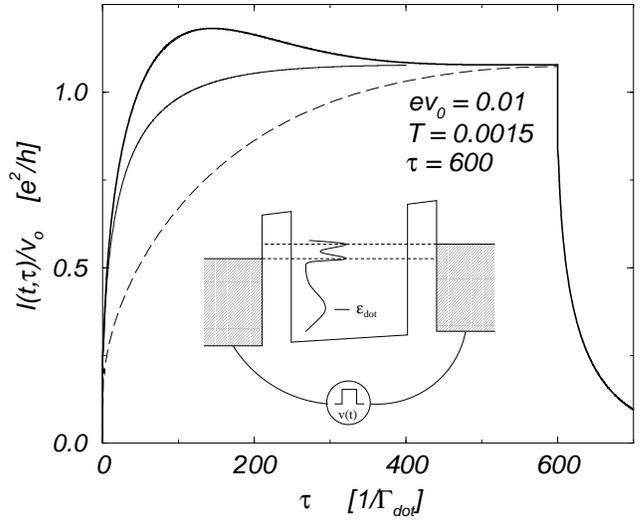}
}
\caption{The instantaneous current $I(t,\tau)$ induced by a bias pulse of
height $ev_0=0.01$ beginning at $t=0$ and ending at $t=\tau=600$ 
(heavy solid line).
The portion for $t > \tau $ is translated back to the origin and
inverted (light solid line) so that the rise time and fall time
can be easily compared.  The area between these curves represents
the excess charge forced through the dot above what would flow
[$G(v_0)\tau$] if the response to $v_0$ were instantaneous.
The value of $v_0$ is small, and these curves represent the
very beginning of the non-linear response regime. As $v_0$ increases further,
the rise time  becomes shorter [$\propto 1/v_0$], while the fall time
remains long.  The dashed line represents the extent to
which the Kondo state would be formed (at zero bias) if the Kondo
coupling were to be suddenly turned on at $t=0$ (see text).
Inset: Schematic of the quantum dot during a pulse.
}
\label{fig:weak}
\end{figure}
We model the quantum dot by a single spin degenerate level 
of energy $\epsilon_{\rm dot}$
coupled to leads through tunnel barriers, as illustrated
schematically in the inset to Fig.~\ref{fig:weak}. The Coulomb charging
energy $U$ prevents the level from being doubly occupied. 
We apply a time-dependent potential across the leads
$v(t)=v_0$ for $0<t < \tau$ and zero otherwise,
so that the energy $E_k$  of an electron in a lead is time
dependent.  Here $k$ represents all lead quantum numbers
other than spin, including the labels, left and right, specifying
the lead in question. 
 Then we have 
$E_k=E_k(t)=\epsilon_k \pm \case{1}{2} ev(t)$, 
the sign depending
on which lead $k$ refers to, where $\epsilon_k$ is the band energy.
The system may be described by the following Anderson
hamiltonian:
\begin{equation}
\sum_\sigma \!
\epsilon_{\rm dot}
n_\sigma +\sum_{k\sigma}\!\left[E_k(t)
n_{k\sigma}
+(V_k c^\dagger_{k\sigma}c_\sigma + {\rm H.c.})\right],
\label{hamiltonian}
\end{equation}
with the constraint that the occupation of the dot cannot
exceed one electron.
Here $c^\dagger_\sigma$ creates an electron of spin $\sigma$
in the quantum dot, with $n_\sigma$ the corresponding
number operator; $c^\dagger_{k\sigma}$ creates an electron in the leads.

The general features of the static equilibrium spectral density when
the dot level $\epsilon$ is sufficiently below the Fermi level 
(taken at zero here)
are well known. There is a broad resonance of 
half-width $\sim$ $\Gamma_{\rm dot}$
\cite{Gnote} at an energy $\sim$ $\epsilon_{\rm dot}$ plus a sharp
temperature sensitive resonance at the Fermi level (the Kondo peak),
characterized by the low energy scale $T_K$ (the Kondo temperature),
$T_K \simeq D' \exp(-\pi|\epsilon_{\rm dot}|/\Gamma_{\rm dot})$,
where $D'$ is a high energy cutoff \cite{dprime}.
Throughout this work energies and temperatures
are given
in units of $\Gamma_{\rm dot}$, and times in units
of $1/\Gamma_{\rm dot}$, with $\hbar=1$.  Specifically, the calculations
in this Letter were made with $\epsilon_{\rm dot}=-2$, which leads to a
Kondo temperature $T_K\sim 0.0025$.  We present results only for
a dot with left-right symmetry.

The  current into the dot  depends on the time $t$ and parametrically
on the pulse length $\tau$, and is given by
\begin{equation}
I_{\rm in}(t,\tau)= 
ie\sum_{k\sigma}V_k \langle c_{k\sigma}^\dagger(t)c_\sigma(t)\rangle
+\mbox{c.c.}
\label{current}
\end{equation}
It may be divided into contributions $I_{\rm left}(t,\tau)$ and
$I_{\rm right}(t,\tau)$ by respectively restricting the $k$ summations
to the appropriate lead. The transport current is then 
 $I(t,\tau)=\case{1}{2}[I_{\rm left}(t,\tau)-I_{\rm right}(t,\tau)]$.
We calculate the Keldysh propagators
corresponding to the angular-bracketed expectation values in
(\ref{current}) for each lead, and hence obtain $I(t,\tau)$.  A more
experimentally accessible quantity, the total charge transported,  
$Q(\tau) = \int_0^\infty dt\,I(t,\tau)$ is also calculated. 
Our calculations  use  the non-crossing approximation (NCA),
which is reliable for temperatures down to
$T < T_K $ \cite{bickers}. Its time-dependent formulation has
been developed elsewhere \cite{TDNCA}.  
The finite bias on the leads is taken into account by introducing a
time-dependent phase in $V_k$ in (\ref{hamiltonian}).

In Fig.~\ref{fig:weak}, we show the response to a long, weak
bias pulse, which illustrates some features that will persist
for still stronger pulses.  The most important feature is that
the rise time of the current is
shorter than the decay time,
leading to a greater charge transport than would have occurred
if the response at each end of the pulse had been instantaneous.

Another comparison can also be made:
how does this time scale for turning on the current 
 compare with how long it takes to form the Kondo state itself?
The latter was
obtained in Ref.~\onlinecite{kondotime}, by calculating the
turn-on of the linear response current when the parameter
$\epsilon_{\rm dot}$ was suddenly switched into the Kondo regime.
 This 
is illustrated by the dashed curve in Fig.~\ref{fig:weak},
 which is the calculation of
Ref.~\onlinecite{kondotime} for the present parameters, with
the abscissa scaled to make the curves coincide at saturation
(large $\tau$ and large $t$, with $t<\tau$).
It is clear that the formation time for the Kondo state is significantly
longer.

A closer look at Fig.~\ref{fig:weak} reveals that 
approximately a third of the initial rise (in both curves) when the
bias is turned on (and the
same fraction of the initial fall when the bias is turned off)
occurs essentially instantaneously on the scale of the graph.
This fast response occurs on
the time scale $t_0\equiv 1/\Gamma_{\rm dot}$.
This represents the part of the conductance arising from the 
fact that the dot level has finite width which overlaps the
Fermi level, and would be present even in the absence of the Kondo peak.

\begin{figure}
\centerline{\epsfxsize=0.47\textwidth
\epsfbox{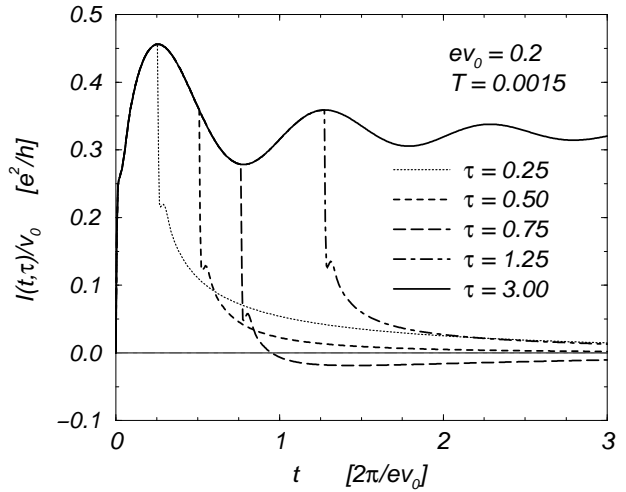}
}
\caption{The instantaneous current $I(t,\tau)$ induced by a strong
pulse (an order of magnitude larger than the width of the
zero bias Kondo peak)
for several pulse lengths $\tau$.  The units of both
$\tau$ and $t$ are both $2\pi/ev_0$, which is almost precisely
the ``period'' of the induced oscillations
(although we see the frequency slightly decrease from $\omega=ev_0$ as 
the bias is lowered towards $T_K$). Note the reversal, for certain pulse lengths,
 of the direction of the instantaneous current
after the end of the pulse.
}
\label{fig:strong}
\end{figure}

The Kondo derived features from switched pulses  can be partially
understood by reference to the {\it steady state} spectral densities
for a dot with a fixed bias equal to $v_0$. Once $ev_0 \gg
\max(T_K,T)$, the Kondo peak becomes split, with a peak separation
associated with the potential  $\pm\case{1}{2}v_0$ of each respective
lead, as discussed by Wingreen  and Meir \cite{theory}. Such spectral
densities for increasing values 
\begin{figure}
\centerline{\epsfxsize=0.48\textwidth
\epsfbox{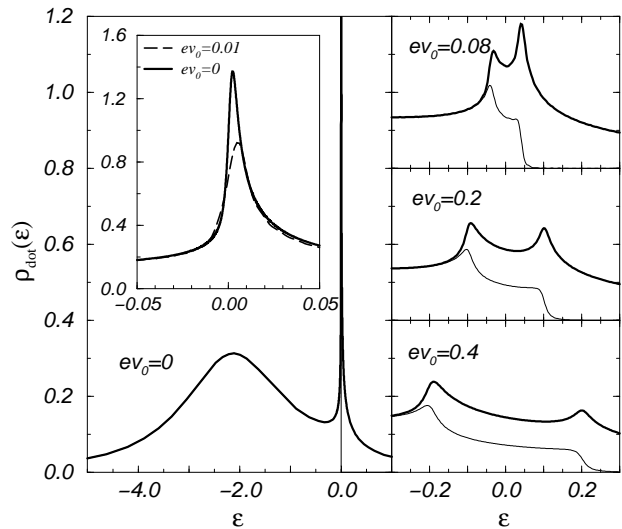}
}
\caption{Steady state spectral densities for different values of the bias.
The left panel shows the whole spectral function for the bias equal to
the $v_0$ of 
Fig.~\ref{fig:weak}, with the Kondo region expanded in the inset;
the equilibrium spectral function is also shown. The three right
panels show just the Kondo region for increasing biases.  The heavy
curves represent the total spectral density, while the light curves
represent the occupied portion.  The oscillations that we find in the
next two figures presumably represent transitions from the lower
Kondo peak in the occupied spectra to the upper peak in the unoccupied
spectra (i.e., the difference between the pairs of curves).
  Note that the vertical scale factors for all four panels
(but, not the inset)
are identical.  $T=0.0015$ everywhere.
}
\label{fig:spectral}
\end{figure}
 \noindent $v_0$ are illustrated in
Fig.~\ref{fig:spectral}.  
This splitting
should provide a sharp excitation  frequency  $\omega \sim ev_0$ for
the dot, with an electron being  excited from the lower peak at energy
$-\case{1}{2}v_0$ to the upper peak at $+\case{1}{2}v_0$.  We should
therefore expect current oscillations of this frequency to be induced
by the rapid turn on of the voltage pulse. For larger $v_0$'s than in
Fig.~\ref{fig:weak}, the frequency $\omega$ provides a fast
characteristic rise time $t_1=\pi/2\omega$, and we may expect that the
rise time depicted in Fig.~\ref{fig:weak}, is a precursor to this
effect. Fig.~\ref{fig:strong}, showing the results for a much larger
$v_0$, not only confirms this interpretation, but shows how
dramatic the current oscillations can be.  In what follows,
we refer to them as {\it split Kondo peak (SKP) oscillations}.

How can one experimentally distinguish SKP
 oscillations from those reported \cite{Jauho94PRB}  
for the resonant level model ($U=0$), which occur as each lead
adjusts to its new chemical potential (NCP)? 
Our calculations show that, unlike the SKP oscillations
which occur robustly at a frequency $\sim$$ev_0$, the NCP oscillations
are sensitive to the details of the exciting pulse. 
First, with the resonant level at the Fermi level to simulate a Kondo
peak, the NCP oscillation frequency is equal to $ev_0$ only in the
 special case that a pulse of magnitude $|ev_0|$ is applied to a
 single lead, with the other held at the original Fermi level.  For
 equal amplitude pulses of $\pm ev_0/2$ to the respective leads, the
 oscillation frequency is, on the other hand, $ev_0/2$. In a more
 general case of asymmetrically applied pulses there are {\it two}
 oscillation frequencies, one associated with the new chemical
 potential of each lead. If in addition, the resonant level is not at
 the original Fermi level, the oscillations are even more complicated,
 with multiple frequency components. 

The NCP oscillations are damped on the time scale $\sim 1/\Gamma_{\rm dot}$.
When this becomes shorter than the oscillation period,
the oscillations become overcritically damped and do not appear.
We find that they only occur for
$ev_0/\Gamma_{\rm dot} \ge 1$, and not for the much smaller
values of this ratio used in the large $U$ examples shown in this paper.
SKP oscillations, on the other hand, occur only 
when  $ev_0/\Gamma_{\rm dot} \ll 1$, and hence not simultaneously
with NCP oscillations.  This explains the robustness of
the oscillations in the Kondo regime.

 SKP oscillations can probably be excited
by other types of applied waveforms, and  a parametrically
driven form of them is probably responsible for the resonances
reported by Schiller and
Hershfield \cite{drivenkondo}.  However, we suspect that 
the well-defined  robust frequency for them
provided by a square pulse, plus 
the lack of one for NCP oscillations, may provide the key for
unambiguous experimental identification.

Fig.~\ref{fig:strong} shows an additional feature---the damping of 
the SKP current oscillations.  This damping presumably arises from the 
fact that the split Kondo peaks in the spectral density have
 widths, giving rise to dephasing.  These widths are substantially
greater than the width of the Kondo peak in equilibrium, as the
application of a 
\noindent bias begins to weaken the Kondo effect.  The
reason for this has been pin-pointed by  Wingreen and Meir \cite{theory},
in a model perturbation theoretic
calculation whose essential feature is that the particle-hole
phase space restriction  from the Pauli principle
in the leads for incoherent spin processes
is eased. For biases $ev_0 \gg T$, this gives a rate enhancement
factor of $\sim$ $ev_0/T$ relative to the zero bias case.
For $T<T_K$, one should replace $T$ by $T_K$ in the above
argument; in this case the width of the Kondo peak(s) would be
proportional to $ev_0$ for $ev_0 \gg T_K$, but would saturate
at $\sim T_K$ for $ev_0 \ll T_K$.  Translating this into
the damping time $t_2$ of the oscillations, gives $t_2
\propto 1/v_0 $ at large biases, but with a saturation  $t_2 \sim 1/T_K$
when $ev_0 < T_K$.  
The expected saturation as the split Kondo
peaks merge removes the oscillations for
$ev_0 < T_K$ ({\it cf.} 
Fig.~\ref{fig:weak}).

Fig.~\ref{fig:strong} also shows the fall off in current after the
voltage pulse has been turned off.  After an initial drop due to the
fast non-Kondo time scale $t_0= 1/\Gamma_{\rm dot}$,
the current falls very slowly. Its decay does not
follow single time scale, but it is clear that the total time required
for the decay is very long. While the initial fall-off after $t_0$ is
bias dependent and relates to the time $t_1$, the long tail has a
characteristic time $t_3 \sim 1/T_{\rm K}$, the longest time in the
problem. This is to be expected: 
 the system is trying to regain an equilibrium where the Kondo
peak is no longer split or broadened.  

Although we only show the results for one temperature, $T=0.0015 \sim
T_K$, we have made calculations for a number of higher and lower
temperatures. The results
\begin{figure}
\centerline{\epsfxsize=0.46\textwidth
\epsfbox{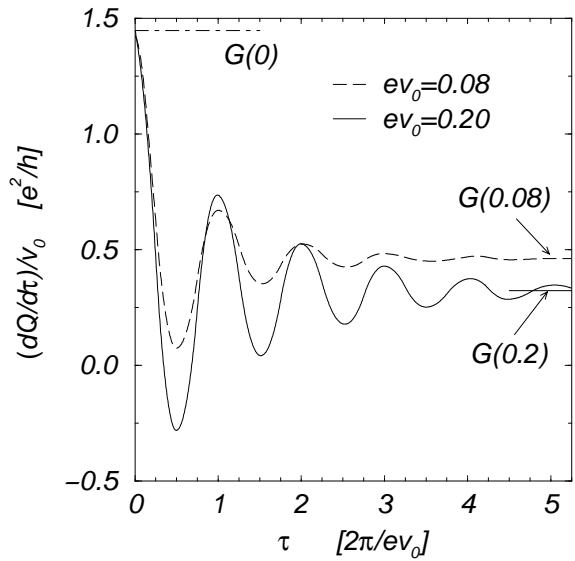}
}
\caption{$1/v_0$ times  the derivative with respect to pulse length $\tau$
 of the total charge $Q$
transported during and after a pulse vs.~$\tau$.
The curves go to
the zero
bias conductance $G(0)$ at $\tau=0$, and each go to the respective
values of the finite bias conductance $G(v_0)$ for large $\tau$.
}
\label{fig:dqdt}
\end{figure}
\noindent  are not very sensitive to temperature when
$T \le T_K$. At higher temperatures, the oscillation amplitude
decreases, and each time $t_1$, $t_2$, or $t_3$
  shortens when $T$ gets large enough to equal the reciprocal
of the time in question ($T\sim 1/t_i$).
 The fall-off time $t_3$ and oscillation decay time $t_2$
 are the first ones
affected. At $T \sim ev_0$, the rise time $t_1$ also shortens. However, the
nonlinear effects persist to temperatures much higher than $T_K$
presumably because of the  logarithmic nature of the Kondo peak.

How can these effects be seen experimentally? We suggest that
the quantity $Q(\tau)$ (the charge
transported through the dot by a single pulse of length $\tau$)
will contain sufficient memory of the charge already transported
at the time $\tau$ when the bias is turned off, to directly reflect
the current oscillations and their damping.  $Q(\tau)$ is proportional
to the average current transported by a series of successive pulses,
provided that the time between pulses is $\gg t_3$; it thus corresponds
to an easily measured quantity.  To restore the features
to the prominence which was there before the time integration,
it is suggested that $dQ(\tau)/d\tau$
be measured. 
  In Fig.~\ref{fig:dqdt} we display our
predicted $dQ(\tau)/d\tau$ plotted vs $\tau$ for a number of values of
bias amplitude $v_0$.  The information about
the frequency and damping of the oscillations, and hence of
the time scales $t_1$ and $t_2$ is preserved. 
Similar  oscillatory structure 
for fixed $\tau$ occurs in plots of $dQ/dv_0$ vs $v_0$.

In conclusion, we have studied the response of a quantum dot
in the Kondo regime to a large pulsed bias voltage across the leads.
We find that the rise time of the instantaneous current
is related to the period of current oscillations that are set up.
These oscillations have a frequency corresponding to the
energy difference between the split Kondo peaks.
The damping rate of these oscillations is related to the widths
of the split Kondo peaks.
 The fall-off
of the current when the pulse is turned off can take
much longer than the rise, with a tail of length approaching $1/T_K$.
These effects may be studied experimentally by measuring
the total charge transported through the dot during and after
a voltage pulse.

The work was supported in part by NSF grants DMR 97-08499 (Rutgers)
and DMR 95-21444 (Rice), DOE Grant DE-FG02-99ER45970 (Rutgers),
and the Robert A. Welch Foundation (Rice).


\begin{thebibliography}{10}
\bibitem{Gnote}
We define $\Gamma_{\rm
  dot}(\epsilon) = 2
  \pi \sum_k |V_k|^2 \delta(\epsilon -\epsilon_k)$, a slowly varying quantity.
 The
  notation $\Gamma_{\rm dot}$ with no energy specified will always
refer to the
  value at the Fermi level.
\bibitem{dprime} For $U=\infty$, $D' \simeq \sqrt{D\Gamma_{\rm dot}/4}$,
where $2D$ is the effective bandwidth.  The calculations here used
a parabolic band of total width $9\Gamma_{\rm dot}$.
\end{thebibliography}
\end{document}